\documentclass[10pt]{article}

\title{Nonlinear integrated optical resonators for optical fibre data recovery}

\author{
	Ivan K. Boikov\textsuperscript{1,*}%
	, Daniel Brunner\textsuperscript{2} %
	and Alfredo De Rossi\textsuperscript{1} \\
	\small\textsuperscript{1}Thales Research \& Technology, Palaiseau Cedex, 91767, France \\
	\small\textsuperscript{2}FEMTO-ST Institute / Optics Department, CNRS \& University Bourgogne\\
	\small Franche-Comte, Besançon Cedex, 25030, France \\
	\small\textsuperscript{*}ivan.boikov@thalesgroup.com
}

\date{}

\usepackage{hyperref}
\usepackage[sorting=none]{biblatex}
\usepackage{graphicx}
\usepackage{amsmath}
\usepackage{dsfont}

\begin{document}

\maketitle

\paragraph{Abstract}
We apply in simulation a reservoir computer based on evanescently coupled GaAs microrings for real-time compensation of a nonlinear distortion of a 50 Gbaud 16-QAM signal with the launch power up to 12 dBm in a standard single-mode optical fibre.
We clearly evidence the crucial role of fast nonlinear response in enabling all-optical signal recovery in real time.
With our system we are able to reduce the signal error rate below the forward error correction limit for a 20~km fibre and 12~dBm launch power.

\section{Introduction}
Optical links are essential for telecommunications.
Their performance in terms of reach and bandwidth, however, is limited by linear and nonlinear distortions present in optical fibres~\cite{amari2017}.
In particular, in short-reach links such as in passive optical networks (PON), a higher launch power would allow for longer reach, more optical network terminals (ONT) and higher-order modulation formats.
However, a nonlinear distortion would be the consequence, whose mitigation in real time is challenging for high modulation bandwidths.

Digital approaches are effective but computationally intensive when running as code on digital hardware using, both, conventional techniques based on digital signal processing (DSP)~\cite{amari2017} and those based on neural networks~\cite{kaneda2021}.
Hence, all-optical methods such as Optical Phase Conjugation (OPC)~\cite{phillips2014} and Phase Conjugated Twin Wave (PCTW)~\cite{liu2013} were proposed.
However, OPC limits flexibility, requiring a precise positioning in a symmetric link~\cite{amari2017}, while PCTW involves the transmission of a twin wave.

One promising alternative is nonlinear distortion mitigation via all-optical neural networks based on the reservoir computing (RC) paradigm.
These simplified recurrent neural networks combine processing dynamical inputs with a much simpler training procedure than generic recurrent neural networks~\cite{jaeger2001}.
Importantly, this simplicity is highly beneficial for hardware implementations~\cite{sande2017}.

Nonlinear transformation is a crucial part of a neural network.
Normally, optical nonlinearities are only observable at the expense of large peak power~\cite{stegeman1988} which can dispute energy efficiency, slow response~\cite{gibbs1979} or large space requirements~\cite{doran1988}.
Semiconductors, and III-V alloys in particular, have been actively investigated in this context because of their relatively large ultrafast Kerr nonlinearity~\cite{aitchison1997} and, more importantly, much larger free carrier-related nonlinear response, typical of semiconductors in passive resonators~\cite{van2002} or amplifiers~\cite{mork1996,nielsen2006}.
Novel concepts such as photonic crystals~\cite{john1987,foresi1997} allowed light confinement within a tiny volume, drastically improving the energy efficiency of all-optical switching~\cite{nozaki2010}, which has long been a major concern.
Finally, interference effects further enhance efficiency without sacrificing speed~\cite{yu2015, bekele2018}.

Such developments unlocked implementations of all-optical RCs, such as those based on a network of splitters~\cite{gooskens2023} and a laser~\cite{argyris2018, estebanez2020, bogris2020} or a nonlinear resonator with an optical feedback~\cite{li2021} that were used for mitigation of linear and nonlinear distortion in optical fibres.
In recent work, we have proposed a novel approach towards scalable RC in integrated photonics based on evanescently coupled nonlinear resonators~\cite{boikov2023}.
This RC allows all-optical real-time processing of coherent high-speed optical signals using nonlinear effects intrinsic to the material platform, for example, III-V heterogeneously integrated on a silicon photonic platform~\cite{bazin2014,constans2019}.
In this letter, we apply this all-optical processing approach to compensate the nonlinear distortion in an optical fibre in real time.

\section{Integrated optical reservoir computer}
A simplified model of a resonator near a waveguide includes the electric field amplitude $a_m$ and free electron density $N_m$:
\begin{equation}
	\begin{cases}
		\dot a_m = -\left( \Gamma_{\rm O}+ \kappa_{\rm in}^2 + \kappa_{\rm out}^2 \right)a_m/2 + \sum\limits_{k \in {\rm near}} \mu a_k + \kappa_{\rm in} s(t) + \dot a_{m,\rm NL} \\
		\dot a_{m, \rm NL} = \left[{\rm i}(\partial_N \omega) N_m - \Gamma_{\rm TPA}^a(|a_m|)/2\right]a_m \\
		\dot N_m = -\Gamma_{\rm C} N_m + \Gamma_{\rm TPA}^N(|a_m|)
	\end{cases},
\end{equation}
where
$\partial_N \omega$ is the free-carrier dispersion coefficient,
$\Gamma_{\rm O}$ is the intrinsic optical loss,
$\kappa_{\rm in/out}$ are the input/output waveguide coupling coefficients,
$\mu$ is the evanescent coupling strength,
$s(t)$ is the input signal,
$\Gamma_{\rm C}$ is the electron recombination rate,
$\Gamma_{\rm TPA}^a(a)$ and $\Gamma_{\rm TPA}^N(a)$ are TPA-related nonlinear functions.
The complete model is given in~\cite{boikov2023}, and our RC is composed of $M$ evanescently coupled resonators with parameters given in Table~\ref{tab:rc}.
\begin{table}[hbtp!]
	\caption{
		Parameters of the reservoir computer.
	}
	\label{tab:rc}
	\begin{center}
		\begin{tabular}[c]{r|l}
			\setlength\intextsep{0pt}
			Geometry & 8\texttimes 3 grid, long side coupled to input \\
			Resonators & GaAs microrings, 5~\textmu m radius \\
			$\mu$ & 75~GHz \\
			$\Gamma_{\rm C}$ & 100~GHz~\cite{moille2015} \\
			$\Gamma_{\rm O}$ & 24~GHz (i.e. $5\cdot 10^4$ Q-factor)\\
			$\kappa_{\rm in}^2$ & $200~{\rm GHz}$ \\
			$\kappa_{\rm out}^2$ & $200~{\rm GHz}$ \\
			TPA constant & 10.2 cm/GW~\cite{combrie2008} \\
			$\partial_N\omega$ & $1.6 \times 10^{-6}$ Hz$\cdot$cm$^3$, using the Drude model
		\end{tabular}
	\end{center}
\end{table}

Reservoirs compute in two steps: a complex nonlinear input expansion into the high-dimensional reservoir space followed by a linear projection through the readout weights to obtain an output.
A rich high-dimensional expansion is likely to improve performance~\cite{skalli2022}.
When operating our system in the linear regime, evanescent coupling creates orthogonal eigenstates with different frequencies referred to as ``supermodes'', even if resonators are identical.
Individual resonators may belong to multiple supermodes and become unique multi-passband filters.
When an input signal is injected into the reservoir, resonators pick up on the the parts of the input that correspond to the resonator's respective frequencies.
As a result, each resonator is excited in a unique and deterministic manner, depending on its associated supermodes and potentially the recent input history.
This way, the input is expanded into a $(2\times M)$-dimensional hyperspace (factor 2 as the reservoir is coherent).
This expansion is further enriched by intrinsic nonlinearities.
Finally, this expansion is linearly projected to produce the output $y(t) = \hat W^{\rm out} \kappa_{\rm out}\vec{a}(t)$, where $\hat W^{\rm out}$ are complex-valued readout weights.

For processing dynamical information, the fading memory requirement of RC~\cite{jaeger2001} implies that reservoir and task timescales should be similar.
The relation between these timescales determines the capability of a RC to keep the recent history of inputs nonlinearly represented in echoes of its internal state, which is crucial for enabling processing of signals that requires information stemming from several timesteps, such as the case of chromatic and nonlinearity compensation in optical communications.
In our RC, there are an optical and an electronic timescale.
The former relies on the photon lifetime, the latter on the free electron lifetime.
In commonly used digital RC subtype called ``echo-state network'' (ESN) the nonlinearity is instantaneous.
As the literature contains many examples of working ESNs, we aim to imitate them.
Since carrier lifetime can be manipulated by controlling the recombination velocity at the surface~\cite{moille2016} or the doping~\cite{moille2017} we choose conditions such that
\[\emph{free~electron~lifetime} < \emph{input~symbol~length} \propto \emph{photon~lifetime}.\]
For such conditions, the fading memory is associated to the photon lifetime.
Determining apriori its optimal length for a task is complicated, however.
We therefore treat the photon lifetime as an optimization parameter, controlled by the coupling strength of microresonators to output waveguides.
The evanescent coupling rate between microresonators defines the input frequency the RC can accept.
However, an excessive coupling rate will harm the dimensionality of the input expansion; a priori we find that matching to the input baud is the best choice~\cite{boikov2023}.

The RC must be trained before the first use and, likely, retrained regularly to account for parameter drift.
Therefore, an efficient training procedure is crucial for practical use.
One good option is ridge regression that requires the knowledge of the internal layer responses to an excitation $\hat A = a_{mk} = a_m(t_k)$, and a corresponding target signal $\vec{Y^{\rm tgt}} = y^{\rm tgt}_{k} = y^{\rm tgt}(t_{k} - \tau^{\rm tgt})$, where $m \in [1\dots M], k \in [1\dots T]$ and $T$ is sufficiently large.
Delaying the target signal with $\tau^{\rm tgt} > 0$ allows the RC to accumulate more relevant information before producing the output, leading to better performance~\cite{boikov2023}.
Finally, the optimal readout weights are
\begin{equation}
	\hat W^{\rm out} = \vec{Y^{\rm tgt}}\hat A\left(\hat A \hat A^* + \beta \hat I\right)^{-1},
	\label{eq:ridge}
\end{equation}
where $^*$ is the conjugate transpose, $\beta$ is a regularization parameter and $\hat I$ is a diagonal matrix.

\section{Short-reach nonlinear optical link}
\begin{figure}[htbp]
	\centering
	\includegraphics{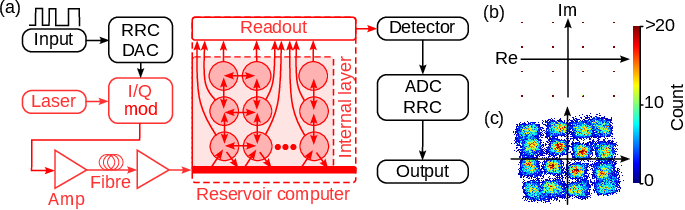}
	\caption{
		(a) Optical link scheme.
		Here, DAC and ADC are digital-to-analog converter and vice-versa, and RRC is root-raised cosine filter.
		Input is passed to the RC via a waveguide (thick red) coupled to a grid of evanescently coupled resonators (circles) that form the internal layer.
		The readout performs weighting and summation of the internal layer responses in the optical domain.
		(b) and (c) show histograms of signal samples on the complex plane before and after a 20~km optical link with 12~dBm launch power that increased the symbol error rate to 0.05.
		The colorbar is shared.
	}
	\label{fig:scheme}
\end{figure}

In this work, we are mostly interested in nonlinear distortion, which could come up in PON.
Compared to C-band links commonly found in literature, PON links are in O-band and are relatively short, meaning dispersion is less important.
However, nonlinearity can be significant, as downstream signals need to be strong enough such that, when split, all optical line terminals still receive sufficient power.
The setup scheme is shown in Figure~\ref{fig:scheme}(a), and parameters given in Table~\ref{tab:param} imitate a PON link before the split.
Here, an optical carrier is modulated with a pseudorandom bit sequence.
The quality of the random number generator is important as otherwise a neural network can learn shortcuts instead of correcting distortion~\cite{eriksson2017}.
Here, Xoshiro256++~\cite{blackman2021} was used.
The propagation inside the standard SMF-28 optical fibre was simulated by solving the nonlinear Schrödinger (NLS) equation~\cite{agrawal2007} with the split-step method.
Noise has been omitted as RC cannot compensate for it~\cite{argyris2018} and is therefore irrelevant to assess RC performance.
\begin{table}[htbp]
	\caption{
		Parameters for simulation of optical signal propagation.
	}
	\centering\begin{tabular}{r | l}
		Modulation & 16-QAM, 50~GBaud\\
		Optical carrier & 1342~nm considered for TDM-PON~\cite{kaneda2021} \\
		Launch power & $6\dots14$~dBm \\
		Pulse shaping & RRC, 0.1 roll-off \\
		Samples per bit & 8 \\
		Optical fibre & up to 70~km SMF-28 \\
		Attenuation & 0.32~dB/km~\cite{smf28} \\
		Dispersion coefficient & 2.4~ps\textsuperscript{2}/km~\cite{smf28} \\
		Kerr coefficient & 1.5~W\textsuperscript{-1}/km~\cite{agrawal2007} \\
	\end{tabular}
	\label{tab:param}
\end{table}

We simulate the RC for 2~\textmu s, which corresponds to $T = 10^5$ symbols, the first half of which is used for training, the other -- for testing.
Consequently, the minimum measurable symbol error rate (SER) is $2\times10^{-5}$.
Simulations have shown that for distortion considered here, $\tau^{\rm tgt}$ equal to a half of the symbol length provides the best performance.

The output power of the RC needs to be considered alongside the SER, as, in practice, noise can be important.
As mentioned above, the RC output is $y(t) = \hat W^{\rm out}\kappa_{\rm out}\vec{a}(t)$.
Since ridge regression does not account for physical constraints, $W^{\rm out}_{i}$ can be any complex number, but there can be constraints specific to a readout implementation.
For example, for an RC with two resonators, the readout can be implemented with a Mach-Zehnder interferometer, which implies $|W^{\rm out}_{1}|^2 + |W^{\rm out}_{2}|^2 = 1$.
Nevertheless, for the task we solve, $\alpha_{\rm read}\hat W^{\rm out}\vec{x}(t)$, where $\alpha_{\rm read} > 0$ is also a solution.
Here, $\alpha_{\rm read}$ can be considered a readout loss; its value depends on both the readout implementation and weights.
As a readout can be implemented in various ways~\cite{bueno2017, shen2017, tait2016, ma2023}, readout loss is a complex topic and is out of scope of this article.
Here, we make a simple assumption that a sum of weighted signals can be computed exactly and there is no amplification at the weighting stage, i.e. all weights are normalized by $\alpha_{\rm read} = 1/{\rm max}_{i}(|W^{\rm out}_{i}|)$.
This readout does not preserve energy, though is expected to suffice for a rough estimate in a typical setting.
Then, we consider the power penalty
\begin{equation}
	10 {\rm log}_{10}\left[ \int {\rm d}t |\alpha_{\rm read}\hat W^{\rm out}\kappa_{\rm out}\vec{a}(t)|^2 \Big/ \int {\rm d}t|s(t)|^2\right]
\end{equation}
as an estimate of the RC optical power use.
The power penalty can be controlled via the regularization parameter $\beta$~\cite{boikov2023}.
Here, we target the power penalty of approximately -20~dB.

In order to demonstrate the contribution of RC's intrinsic nonlinearities, we compare performance of the RC with a linear tapped filter
\begin{equation}
	y_{\rm filter}(t) = \sum\limits_{l = -L}^{L} w_{l}s\left(t + l\frac{\Delta t}{4}\right),
\end{equation}
where
$\Delta t$ is the input symbol length and
$w_l$ are complex-valued weights of a tap trained with ridge regression using Eq.~\ref{eq:ridge}, where $\hat A$ is replaced by
\begin{equation}
	\hat S = s_{lk} = s\left[t_k - (l-1-L)\Delta t/4\right].
\end{equation}
Here, we chose $L = 12$ to obtain 25 taps for a fair comparison with the RC that has 24 resonators, although $L > 4$ does not improve the equalization significantly.
We do not consider the RC in the linear regime as it produces almost the same result as the tapped filter.

\begin{figure}[htbp]
	\centering
	\includegraphics{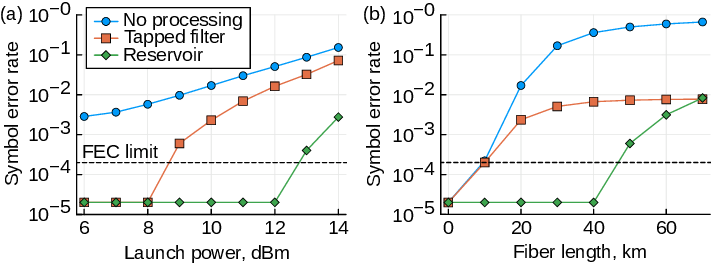}
	\caption{
		Recovery performance of reservoir compared to a tapped filter.
		The legend is shared.
		In (a) the fibre length is 20~km, in (b) the launch power is 10~dBm, and the Forward Error Correction (FEC) limit is 0.2\texttimes 10\textsuperscript{-3}.
	}
	\label{fig:nlcheq-curves}
\end{figure}

\begin{figure}[htbp]
	\centering
	\includegraphics{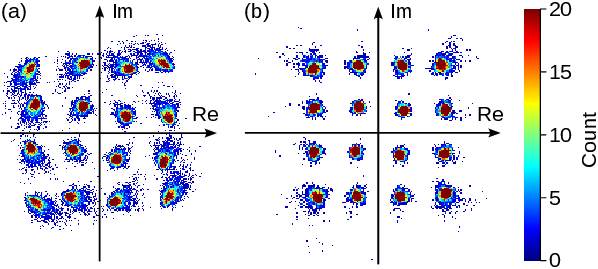}
	\caption{
		Histograms of samples of an optical signal after 20~km of optical fibre with 12~dBm launch power and recovered with (a) a tapped filter and (b) the reservoir computer.
		The colorbar is shared.
		In (a) the SER is 1.6\texttimes10\textsuperscript{-2}, in (b) 2\texttimes10\textsuperscript{-5}.
		The unprocessed signal is in Figure~\ref{fig:scheme}(c)
	}
	\label{fig:nlcheq-hist}
\end{figure}

The equalization result is shown in Figure~\ref{fig:nlcheq-curves} and Figure~\ref{fig:nlcheq-hist}.
The RC input power was optimized: in Figure~\ref{fig:nlcheq-curves}(a) it increased proportionally to the launch power from 10~mW at 6~dBm to 30~mW at 14~dBm, and in Figure~\ref{fig:nlcheq-curves}(b) it was fixed at 25~mW.

\begin{figure}[htbp]
	\centering
	\includegraphics{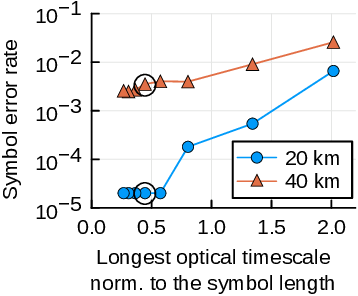}
	\caption{
		The impact of excess memory on equalization at 12~dBm launch power.
		The optical timescale is varied by coupling to input and output waveguides.
		Black circles correspond to parameters in Table~\ref{tab:rc}.
		The legend shows the optical fibre length.
	}
	\label{fig:nlcheq-vs-photon}
\end{figure}

As mentioned above, the optical timescale is an optimization parameter.
It can be increased with a smaller intrinsic Q-factor of resonators, or their weaker coupling to input and output waveguides and vice-versa.
Figure~\ref{fig:nlcheq-vs-photon} shows a sweep of the latter.
The problem favors a shorter optical timescale, although the effect is less prominent for a longer fibre.
For a shorter fibre the performance can be further improved relative to default parameters.
However, this would require a lower Q-factor, which will reduce efficiency, or stronger coupling to waveguides, which may pose a design challenge.

\section{Discussion}
We observed that an excess photon lifetime degrades the performance; its optimal value was smaller than the symbol length.
A possible explanation could be that dispersion was too weak to require a long memory.
A short photon lifetime implies a large bandwidth of supermodes, which limits scalability of RC dimensionality in the linear regime~\cite{boikov2023}, here, roughly 8 degrees of freedom, with free-carrier dispersion improving to roughly 12.
It is possible that other degrees of freedom can still contribute to performance, but regularization prevents them due to their low amplitude.

Due to nonlinearity in the fibre, sidebands appear in the optical spectrum.
In principle, sidebands also carry information that the RC can utilize.
With that motivation, we have increased the evanescent coupling strength to allow some supermodes to overlap with sidebands.
However, sidebands are typically weaker than the main band and these supermodes were weakly excited.
A higher input power would lead to stronger sidebands, but the same is true for the main band, and supermodes there are at risk of losing consistency~\cite{boikov2023}.
One possibility is utilizing smaller features of RC responses by reducing noise and regularization, as seen in Figure~\ref{fig:nlcheq-hist}(c).

We have assumed a free electron recombination rate of 100~GHz, which can be a challenge from the point of view of technology.
Roughly, silicon photonics can provide up to 10~GHz, but III-V materials allow for a much higher ceiling~\cite{constans2019}.
In particular, GaAs can easily fit the requirements and, if necessary, can be slowed down~\cite{moille2015}.
In this work we assumed GaAs microrings, but photonic crystals have the potential to reduce the input power requirement and further accelerate electron recombination~\cite{constans2019}.
Since the nonlinear timescale was faster then the input timescale, the result is expected to be extendable towards materials with the Kerr effect as the dominant nonlinearity such as silicon nitride.

Due to a lack of enforced structure and control of the internal layer, reservoirs are often claimed to be resilient to parameter deviation arising, for example, due to fabrication tolerances.
The computing performance of our RC is largely defined by its spectral characteristics and supermodes in particular.
However, supermodes will only form if the resonator's frequencies are not too different.
If the standard deviation of resonance frequencies exceeds half of the evanescent coupling strength, the dimensionality of the input expansion starts to deteriorate~\cite{boikov2023}.
Using the thermooptic effect can mitigate this problem to an extent, however, a close proximity of resonators will lead to a strong thermal crosstalk.

\begin{figure}[htbp]
	\centering
	\includegraphics{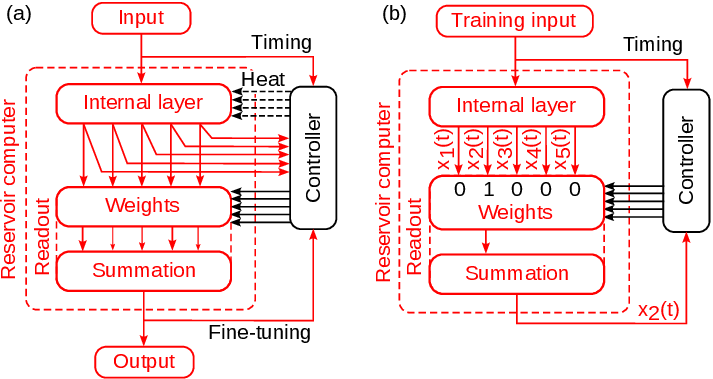}
	\caption{
		(a)~An optimistic use of the proposed RC with, for clarity, 5 resonators.
		The controller aligns resonances of resonators by using heaters.
		(b)~An alternative method for recording internal layer responses.
		Here, the controller records the second resonator response.
		Details in text.
	}
	\label{fig:training}
\end{figure}

In simulation, carrying out the training procedure with ridge regression is straightforward, as the RC state is fully observable, and the training target is available as the distorted signal can be prepared offline.
In practice, the transmitter and the receiver can agree on a training initiation protocol and the training signal to be transmitted; no other data is sent during that time.
Then, a distorted signal excites the internal layer, and the controller records its responses.
Figure~\ref{fig:training}(a) shows a possible operation of such an RC in practice.
This approach closely follows the RC principle, where all neurons are recorded simultaneously.
However, each neuron will require a separate RF channel with a coherent detector, which is costly and a design challenge even for a modest RC.
An alternative approach is shown in Figure~\ref{fig:training}(b).
During the training, the transmitter sends the same signal multiple times.
Each time, the controller sets all weights to zero except one set to unity.
This way, only one response passes through the readout unchanged and is then recorded by the controller.
The controller can record all responses reusing the output port by cycling through all weights.
Here, speed of training is traded for simplicity of design.
If implementing ridge regression poses a challenge, black-box optimization concepts such as reinforcement learning~\cite{bueno2017} or evolutionary algorithms~\cite{buckley2023} can be a viable alternative.

Ridge regression might take a considerable amount of time, during which the optical link cannot be used.
However, if there is only a minor parameter drift, the controller can account for them online using black-box optimization methods such as gradient descent or methods mentioned above.

\section{Conclusions}
An integrated all-optical reservoir computer based on evanescently coupled resonators is viable for compensating nonlinear distortion in an optical fibre.
Here, the reservoir recovered the distortion of a 50~Gbaud signal after a short-reach link in O-band with 20~km length and 12~dBm of optical power injected, a case relevant for the next generation of passive optical networks.
The computing was performed in real time while using a few tens of milliwatt of optical power.

We found that a low photon lifetime, although harmful for dimensionality, positively impacts performance in this task.
The reservoir also poses a few challenges with regards to technology and design, such as engineering of the free electron lifetime, which has been demonstrated to be possible.

\section*{Acknowledgements}
The authors thank David Bitauld, François Duport, Peter Bienstman, Serge Massar and Sergei Turitsyn for fruitful discussions.
A part of this work was supported by European Union through the Marie Skłodowska-Curie Innovative Training Networks action, POST-DIGITAL project number 860360.

\printbibliography
\end{document}